\documentclass[]{spie}  

\usepackage{amsmath,amsfonts,amssymb}
\usepackage{graphicx}
\usepackage[colorlinks=true, allcolors=blue]{hyperref}

\usepackage{siunitx}

\usepackage{caption}
\usepackage{subcaption}

\usepackage{color}

\title{Tunable interferometers on a flexible polymer chip}

\author[a]{James A. Grieve}
\author[a]{Chengcan Li}
\author[a]{Kian Fong Ng}
\author[b]{Yi Wei Ho}
\author[b,c]{Jos\'{e} Viana-Gomes}
\author[a,c]{Alexander Ling}

\affil[a]{Centre for Quantum Technologies, National University of Singapore, 3 Science Drive 2, 117543 Singapore}
\affil[b]{Centre for Advanced 2D Materials, National University of Singapore, 6 Science Drive 2, 117546 Singapore}
\affil[c]{Department of Physics, National University of Singapore, 2 Science Drive 3, 117551 Singapore}

\authorinfo{Further author information:\\J.A.G.: E-mail: james.grieve@nus.edu.sg\\  A.L.: E-mail: cqtalej@nus.edu.sg}

\begin{document} 
\maketitle

\begin{abstract}
To facilitate the implementation of large scale photonic quantum walks, we have developed a polymer waveguide platform capable of robust, polarization insensitive single mode guiding over a broad range of visible and near-infrared wavelengths. These devices have considerable elasticity, which we exploit to enable tuning of optical behaviour by precise mechanical deformations. In this work, we investigate pairs of beamsplitters arranged as interferometers. These systems demonstrate stable operation over a wide range of phases and reflectivities. We discuss device performance, and present an outlook on flexible polymer chips supporting large, reconfigurable optical circuits.
\end{abstract}


\section{INTRODUCTION}
\label{sec:intro}  

Flexible waveguide platforms based on elastomeric substrates have been proposed as solutions to a wide range of problems in integrated photonics~\cite{Hu2013}, with notable applications demonstrated in the areas of optical interconnects~\cite{Missinne2014} and tunable photonic devices such as interferometers and resonators~\cite{Chen2012}. The availability of inexpensive and high quality optical-grade polymers combined with relatively simple handling requirements make such devices an attractive target for both industrial and academic research.

In an extension of our previous work on broadband tunable beamsplitters~\cite{Grieve2017} and coupled waveguide arrays~\cite{Grieve2018}, we are investigating networks of cascaded beamsplitters arranged to form a discrete time, discrete space quantum walk. To evaluate the stability and suitability of this platform, we begin with pairs of splitters arranged as interferometers. We envisage small scale devices of this type finding use in optical sensing~\cite{Odeh2017}, while larger systems may be useful in implementations of photonic simulator devices~\cite{Tillmann2013,Crespi2013,Grafe2016}.

\section{METHODS}
\label{sec:methods}

\begin{figure}[bt]
\centering
\includegraphics[width=10cm]{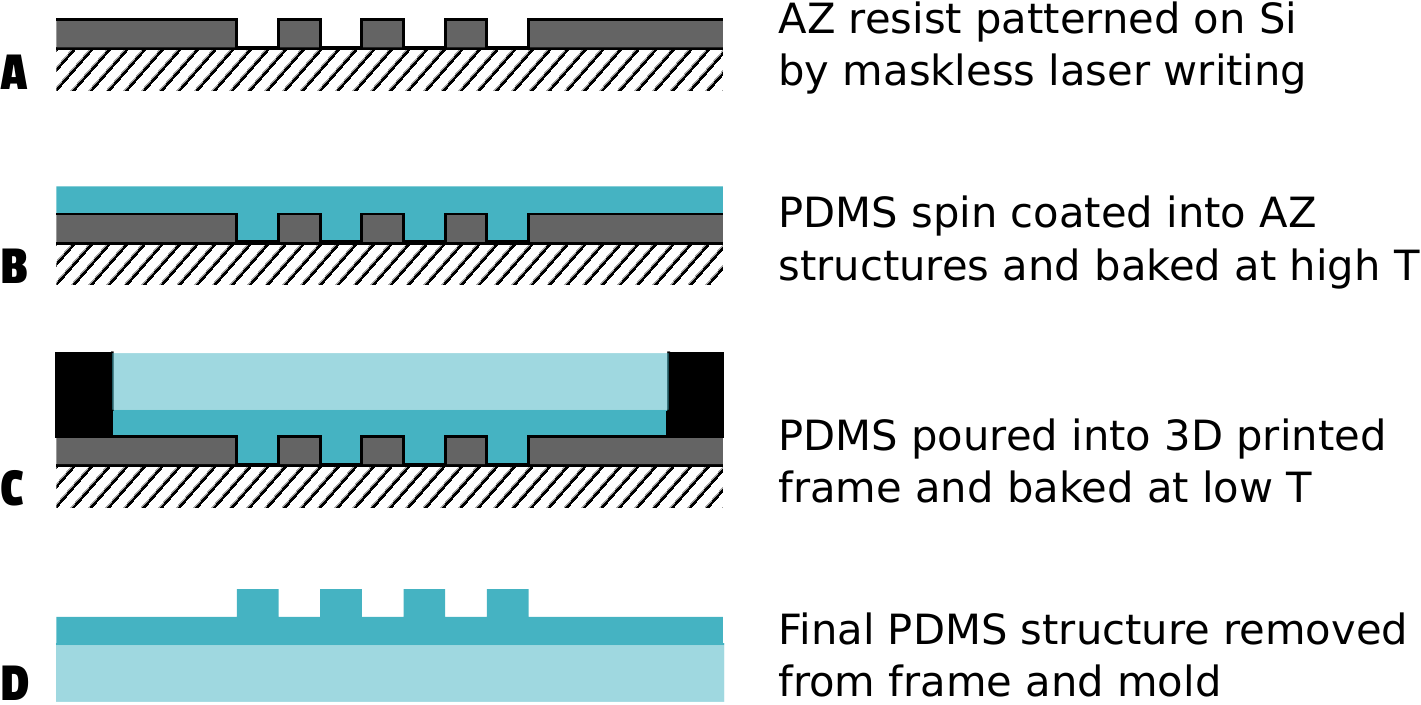}
\vspace{0.4cm}
\caption{\label{fig:fabrication}A simplified diagram showing the fabrication process for single mode polydimethylsiloxane (PDMS) waveguide devices. \emph{Reproduced with permission from Appl Phys Let. 111(21) (2017).} }
\end{figure}

\begin{figure}[b]
\centering
\begin{subfigure}[t]{7cm}
  \centering
  \includegraphics[height=4cm]{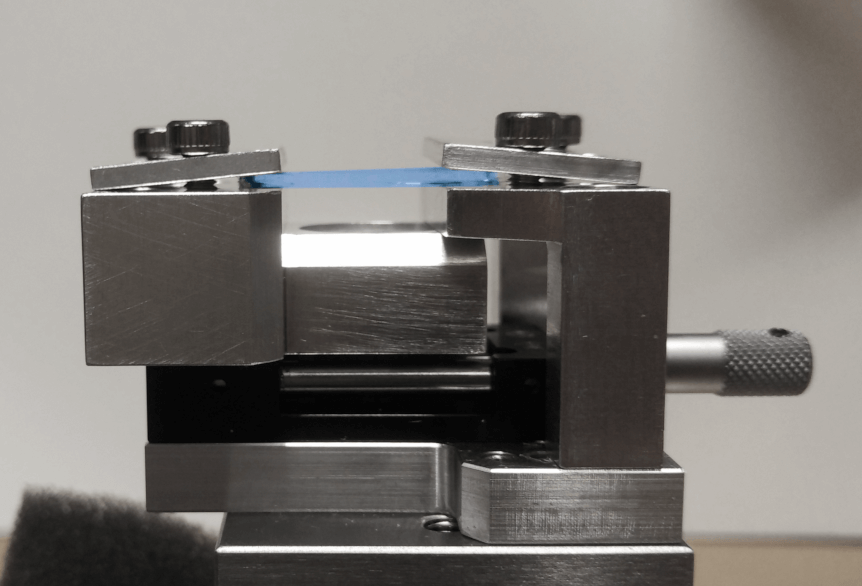}
  \vspace{0.5cm}
  \caption{\label{fig:jig:photo}}
\end{subfigure}
\quad
\begin{subfigure}[t]{6cm}
  \centering
  \includegraphics[height=5cm]{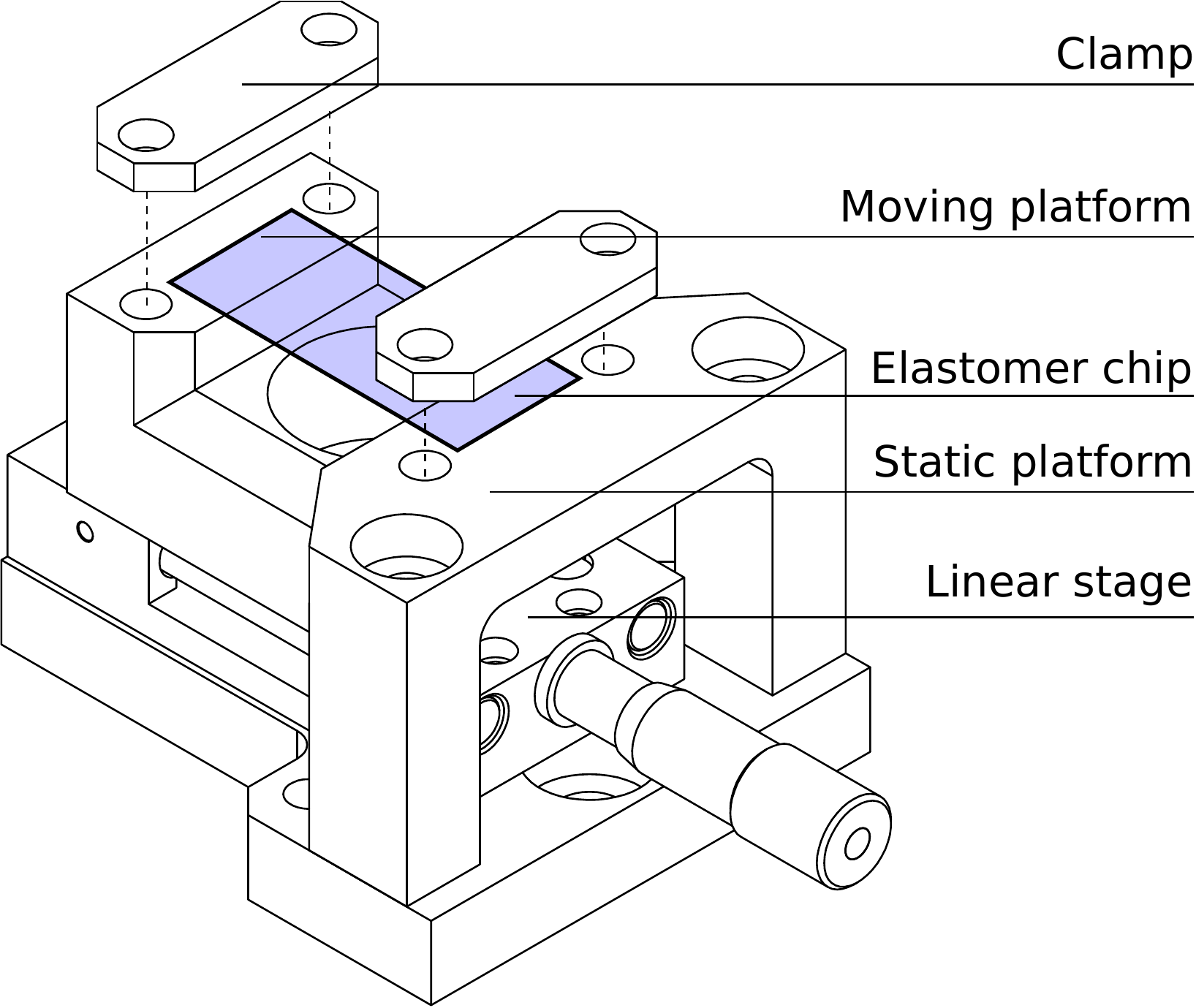}
  \caption{\label{fig:jig:diagram}}
\end{subfigure}
\caption{\label{fig:jig}The elastomer chip is mounted on a custom stretching jig driven by a miniature linear stage (Newport Corporation), which enables the controlled application of strain orthogonal to the optical axis. In (a) the chip is shown mounted, with false color applied to increase visibility of the transparent material. An exploded schematic is shown in (b).}
\end{figure}

Waveguides are fabricated in a two-layer monolithic elastomeric platform based on polydimethylsiloxane (PDMS) (Sylgard 184, Dow Corning), using a technique adapted from Kee et~al.\cite{Kee2009} and reported previously\cite{Grieve2017}. A negative pattern is first defined in a \SI{1.7}{\micro\meter} layer of photoresist (AZ1512HS, AZ Electronic Materials) on a silicon wafer by direct laser writing, which serves as a mold for the final device and can be reused several times. Liquid PDMS is introduced to the mold and spun to a thickness of approximately \SI{5}{\micro\meter}. After curing for \SI{1}{\hour} at \SI{150}{\celsius}, a second, thicker layer of PDMS is poured onto the stack and cured at \SI{70}{\celsius}, with the lower temperature giving rise to the required refractive index contrast. The device is peeled from the mold and mounted on a custom stretching jig (see Figure~\ref{fig:jig}). A simplified schematic of this process is shown in Figure~\ref{fig:fabrication}.

Waveguides produced in this manner show robust single mode guiding from \SIrange{450}{850}{\nano\meter}~\cite{Grieve2017}, with propagation losses typically lower than \SI{0.1}{\decibel\per\milli\meter}. The devices are also relatively insensitive to polarization, with low birefringence and negligible depolarization (a degree of polarization of \SI{0.9992(4)}{\per\centi\meter} and \SI{0.9979(13)}{\per\centi\meter} for transmitted horizontal and vertical input states respectively).

\begin{figure}[hbt]
\centering
\begin{subfigure}[t]{7cm}
  \centering
  \includegraphics[width=7cm]{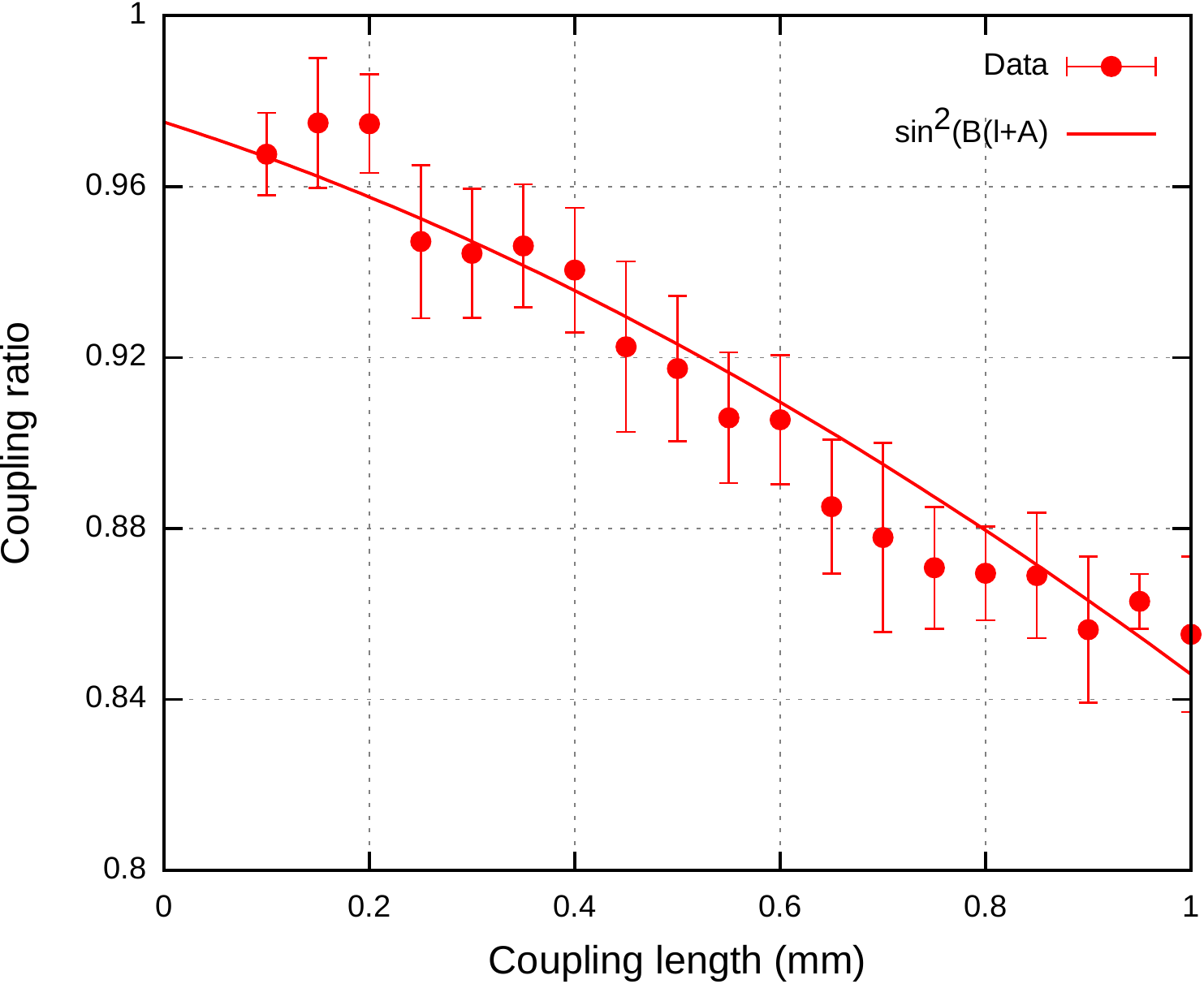}
  \caption{\label{fig:1a}}
\end{subfigure}
\quad
\begin{subfigure}[t]{7cm}
  \centering
  \includegraphics[width=7cm]{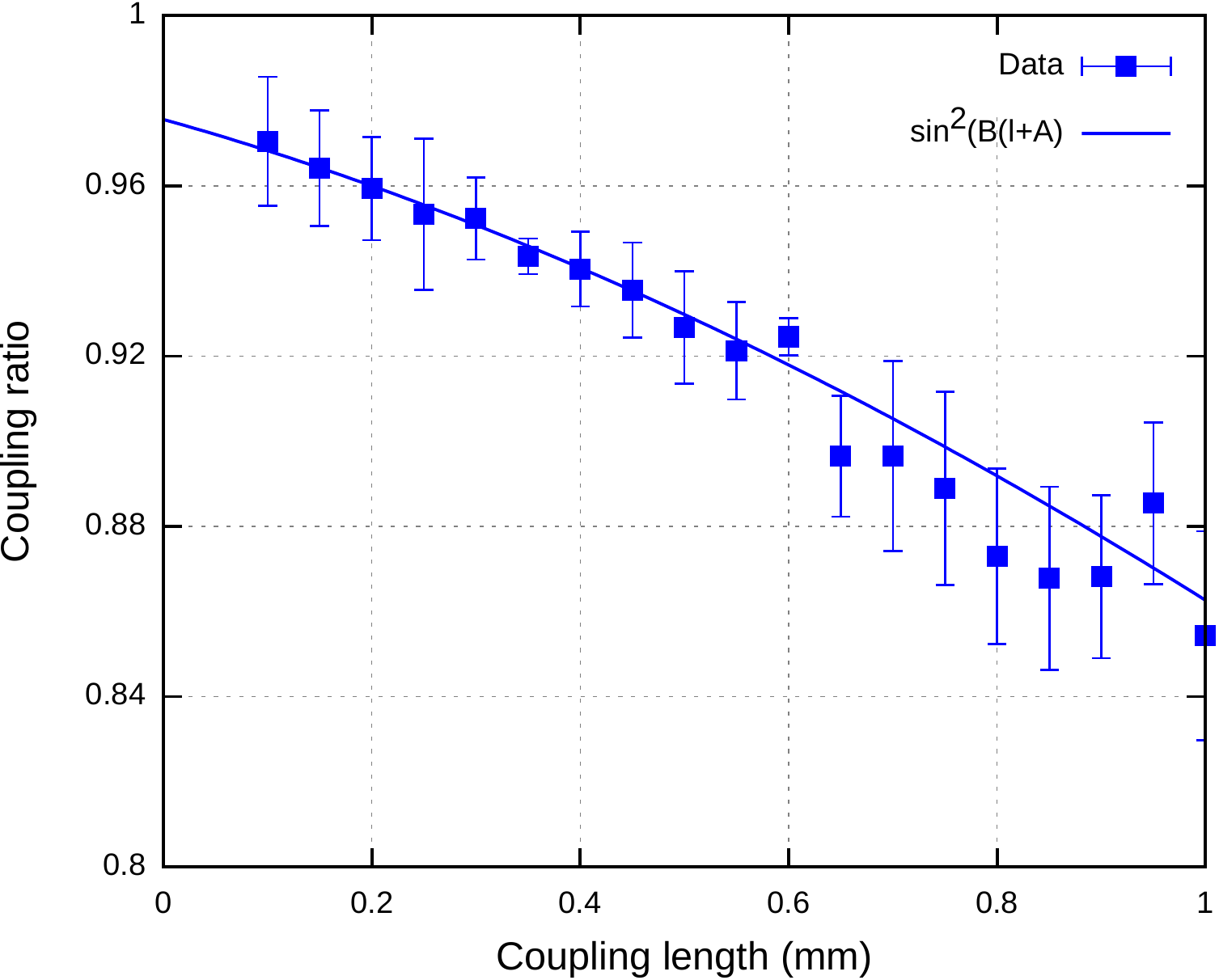}
  \caption{\label{fig:1b}}
\end{subfigure}
\caption{\label{fig:variablelengths}Splitting ratio of directional couplers as a function of their designed coupling length. Two sets of couplers were fabricated and characterized, with consistent results. The design of the devices is such that evanescent coupling is significant during the fan-in and fan-out regions, so the \emph{effective} coupling length is greater than specified. In this case, nearly full coupling is observed for extremely short `design' lengths, and reduces with increasing length.}
\end{figure}

We define directional couplers in the device with an input pitch of \SI{50}{\micro\meter}, with \SI{4}{\milli\meter} fan-in and fan-out regions and varying coupling lengths. In the coupling region, waveguides are separated by \SI{5.7}{\micro\meter} (approximately equal to their width). Due to the very large bend radius of this design, the waveguides are evanescently coupled far outside of the designed ``coupling region'', so the \emph{effective} coupling length is typically observed to be much greater than the designed value. A survey of a variety of such coupling lengths is shown in Figure~\ref{fig:variablelengths}. The coupling ratio (defined as the ratio of power in the uncoupled arm to the power in both uncoupled and input arms) follows the expected $\sin^2$ scaling~\cite{Lifante2003a}, allowing us to design devices with targeted coupling.

As these directional couplers will be operated with applied strain in the direction perpendicular to the optical axis, we concentrate on devices with ``over coupled'' behaviour, i.e. with full or nearly full coupling from the input to the uncoupled arm when unstretched. From our previous work~\cite{Grieve2017}, we expect to see reduction in this ratio as the devices are tuned mechanically.

The behaviour of a single directional coupler in this elastomeric platform has been described previously~\cite{Grieve2017}, and can be modeled effectively via coupled mode theory~\cite{Lifante2003a}. We can capture this behaviour using the ratio of transmitted power to total power, $T_{BS}$:

\begin{equation}
\label{eqn:tbs}
T_{BS} = \frac{P_{T}}{P_{T}+P_{R}},
\end{equation}

\noindent where $P_{T}$ and $P_{R}$ are the transmitted and reflected power respectively. These parameters are sensitive to mechanical deformation of the chip, and can be tuned by stretching the device. Where two such directional couplers are connected in series to form an interferometer, the transmitted and reflected power of the system is a function of the transmission of the individual directional couplers and a phase term ($\phi$) related to the potentially unequal path length of the waveguides between the couplers. Using the same naming convention, we can define transmitted and reflected powers $P_{T,MZI}$ and $P_{R,MZI}$ and reduce the behaviour of a single interferometer to its transmission (defined as for the single directional coupler),

\begin{equation}
\label{eqn:tmzi}
T_{MZI} = \frac{P_{T,MZI}}{P_{T,MZI}+P_{R,MZI}} = (2\cos\phi + 2) T_{BS} (T_{BS}-1)+1.
\end{equation}

From this expression the overall scaling behaviour of the device is clear: we anticipate a cosine variation in terms of the relative phase of the two paths, with a ``visibility'' which is maximised for the case of $T_{BS} = 0.5$.

\section{RESULTS AND DISCUSSION}
\label{sec:results}

Several devices were designed and fabricated on the same chip following the process detailed in the previous section. With the chip mounted as shown in Figure~\ref{fig:jig}, \SI{632}{\nano\meter} light from a HeNe laser was coupled to one arm at the input face using edge-coupling with a single mode optical fiber. The output face of the chip was imaged onto a calibrated CCD (Chameleon3, Flir) where the relative power of the output modes could be calculated by integrating pixel values within appropriately defined regions of interest. The relative transmission of the interferometer (defined as in Equation~\ref{eqn:tmzi}) was recorded for a variety of applied strains.

The first observation is qualitative in nature. One possible concern with the adoption of a pliable elastomeric chip material relates to the temporal stability of the resulting device. In our previous work~\cite{Grieve2017}, directional couplers were evaluated in isolation and were found to be relatively insensitive to mechanical disturbance, enabling tuning by relatively large applied strain (up to 30\,\%). In the work described here, devices are expected to be significantly more sensitive to mechanical vibration owing to the phase sensitivity of the interferometric configuration. However, mounted chips did not show the characteristic ``blinking'' commonly associated with free-space implementations without active stabilization.

\begin{figure}[hbt]
\centering
\begin{subfigure}[h]{7cm}
  \centering
  \vspace{0.3cm}
  \includegraphics[width=7cm]{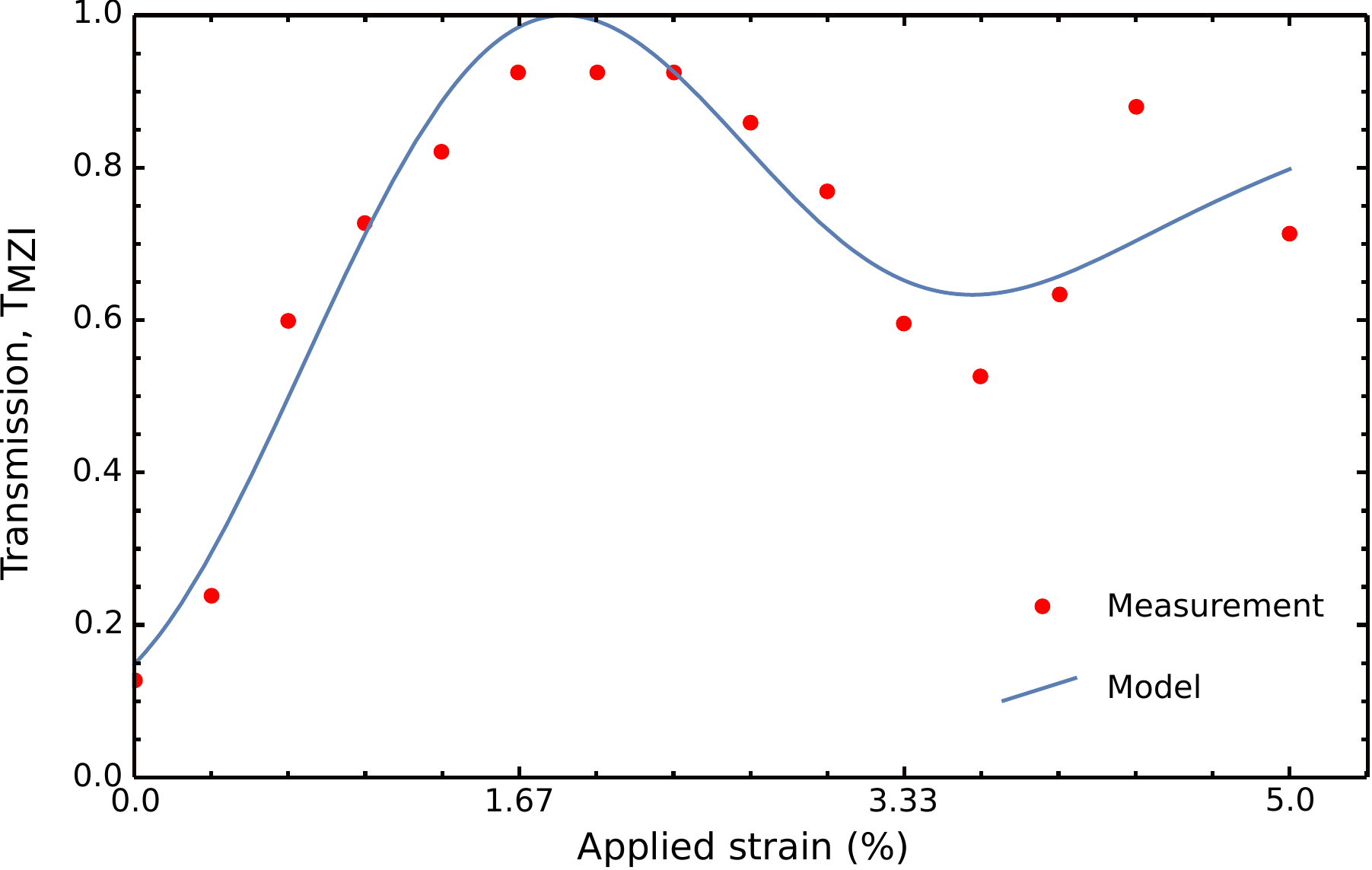}
  \vspace{0.1cm}
  \caption{\label{fig:mzidata:2d}}
\end{subfigure}
\quad
\begin{subfigure}[h]{8cm}
  \centering
  \includegraphics[width=8cm]{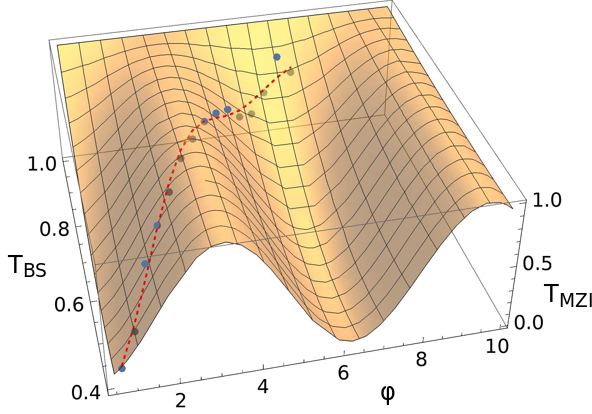}
  \caption{\label{fig:mzidata:3d}}
\end{subfigure}
\caption{\label{fig:mzidata}Experimentally measured transmission for an interferometer undergoing distortion. (a) Measured device transmission ($T_{MZI}$) plotted against applied strain, fit using Equation~\ref{eqn:tmzi}, assuming a $\sin^2$ relationship for the transmission of the individual couplers and a linear relationship between applied strain and the unbalanced phase term. (b) Data points are visualized on a surface of possible values given by a range of beamsplitter transmissions ($T_{BS}$) and relative phases ($\phi$). Data is plotted according to the fit in (a), with the dashed line representing the model and the spheres measured transmission. Points above the surface appear blue, while those below appear black.}
\end{figure}

Due to the small size of the device we are not able to independently control the strain applied to the directional couplers, and we are unable to apply controlled distortion to the paths inside the interferometer. Nevertheless, some interesting observations can be made from applying global strain. Figure~\ref{fig:mzidata} shows the transmission of a representative interferometer based on a pair of directional couplers with design coupling length \SI{100}{\micro\meter} and a separation of \SI{4.7}{\micro\meter} (design $T = 0.5$) as a function of applied strain. The approximately \SI{30}{\milli\meter} chip was stretched by up to \SI{1.6}{\milli\meter} yielding a strain of approximately 5.3\,\%. The qualitative behaviour of the device is markedly different from the single directional couplers reported previously~\cite{Grieve2017}, with transmission initially increasing before decreasing to approximately 0.5 and increasing again. It is not possible to reproduce this behaviour using a model that ignores the contribution of relative phase in the interferometer arms, indicating that some asymmetric distortion of the chip must be present.

We fit the data to the model described in Equation~\ref{eqn:tmzi}, further assuming a $\sin^2$ relationship between the individual splitter behaviour $T_{BS}$ and a linear relationship between the global strain and the phase term $\phi$. To better visualize this model, the data is also represented on a 3D surface following Equation~\ref{eqn:tmzi}. We note that although it is difficult to precisely infer phase or splitting ratio from this data, the measurements do follow a plausible path over this surface.

\begin{figure}[hbt]
\centering
\begin{subfigure}[h]{5cm}
  \centering
  \includegraphics[width=3cm]{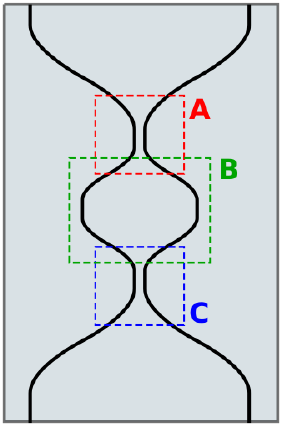}
  \caption{\label{fig:stretchdiagram:1}}
\end{subfigure}
\begin{subfigure}[h]{5cm}
  \centering
  \includegraphics[width=4.3cm]{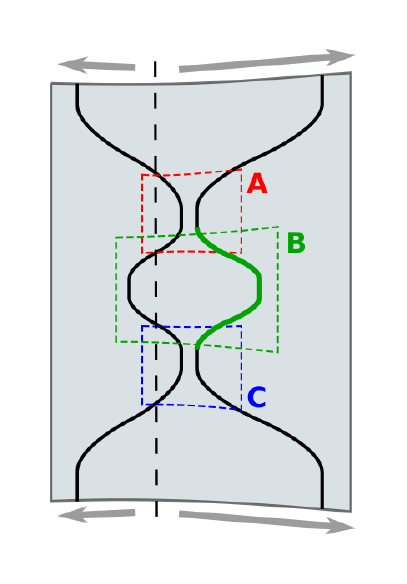}
  \caption{\label{fig:stretchdiagram:2}}
\end{subfigure}
\caption{\label{fig:stretchdiagram}Stylized diagram showing distortion of the cascaded beamsplitter device. (a) The device in the initial state. Coupling regions (denoted A and C) are symmetric, with the internal arms (contained in the region denoted B) giving equal path lengths. (b) The device in a distorted state. Couplers (regions A and C) are not sensitive to asymmetric distortion, so applied strain reduces their transmission as intended. As the device is not precisely centred in the chip, the stretching is slightly asymmetric (a hypothetical centre line is shown, dashed black) giving rise to an increase in the path length of one arm (shown in green) and an associated change in the relative phase as a consequence of Poisson's ratio. }
\end{figure}

Although the device is symmetric about the optical axis, an additional unbalanced phase is seen in response to the global strain. We trace this effect to the location of the device within the chip. As devices are not in general centered between the clamping points, the effect of Poisson's ratio on the dimensions of the distorted chip will lead to a slight variation in the length of the two arms of the device. Figure~\ref{fig:stretchdiagram} shows a stylized diagram of this mechanism, with dashed regions (A,B,C) denoting the first directional coupler, internal paths and second directional coupler respectively. If the device is not precisely centered with respect to the edges of the chip as held by the clamping mechanism (Figure~\ref{fig:jig}), the effect of Poisson's ratio will be to shrink the chip slightly parallel to the optical axis of the imaging system. This will cause one arm of the device to experience a small increase in its length relative to the other. While to first order this is not expected to influence the behaviour of the directional couplers, our device is highly sensitive to changes in the internal path lengths (region B, Figure~\ref{fig:stretchdiagram}(b)).\\

\noindent These observations highlight the limitations of global distortion on such a system. While still potentially useful in applying a correcting strain to all directional couplers in a device, it is clear that larger optical circuits involving many directional couplers will need to include mechanisms for trimming the relative path length between couplers. To this end, work is currently ongoing to develop solutions for applying out-of-plane distortion to the devices, as well as measures to reduce the cross-talk between neighbouring actuation points.

The liquid pre-polymer stage of the PDMS waveguide fabrication method may also allow the precise embedding of piezoelectric transducers inside the bulk of such devices, without the need for complex multi-material lithography steps. With the addition of this capability, highly sensitive strain fields could be manipulated, enabling the precise tuning of individual waveguide elements. While much remains to be done in optimizing this platform, we believe it represents an interesting and promising avenue for the exploration of low cost mechanically tunable photonic systems.

\acknowledgments 
 
The authors would like to thank Filip Auksztol for helpful discussions. This work was supported by the Singapore Ministry of Education Academic Research Fund Tier 3 (Grant No. MOE2012-T3-1-009) and the National Research Foundation, Prime Minister’s Office, Singapore under its Research Centres of Excellence programme.

\bibliography{report} 
\bibliographystyle{spiebib} 

\end{document}